\documentclass[preprint,aps,nofootinbib,superscriptaddress,11pt]{revtex4-1}
\usepackage{amssymb,mathrsfs,physics,amsthm,amssymb,amsfonts,amsmath}

\usepackage{upgreek}

\usepackage{import}

\usepackage{hyperref}
\hypersetup{
    plainpages=false,       
    unicode=false,          
    pdftoolbar=true,        
    pdfmenubar=true,        
    pdffitwindow=false,     
    pdfstartview={FitH},    
    pdfnewwindow=true,      
    colorlinks=true,        
    linkcolor=Blue,         
    citecolor=Blue,        
    filecolor=magenta,      
    urlcolor=blue           
}
\usepackage[usenames, dvipsnames]{xcolor}

\newcommand\id{\leavevmode\hbox{\small1\kern-3.3pt\normalsize1}}

\usepackage{enumitem}

\usepackage[normalem]{ulem}

\theoremstyle{definition}

\usepackage{chngcntr}
\usepackage{apptools}
\AtAppendix{\counterwithin{theorem}{section}}

\usepackage{graphicx}
\graphicspath{ {images/} }

\usepackage{caption}

\usepackage{tabularx}

\usepackage{bbm}

\usepackage{tikz}
\usetikzlibrary{arrows}

\newcommand{\cps}{\mathrel{\ooalign{$\nearrow$\cr \kern-0pt$\nwarrow$}}}

\usepackage{CJKutf8}

\usepackage{versions}
\excludeversion{cmt}

\usepackage{xparse}

\NewDocumentCommand\LH{mo}{%
  \IfNoValueTF{#2}
   {\mathcal{L}(\mathcal{H}^{#1})}
   {\mathcal{L}(\mathcal{H}^{#1},\mathcal{H}^{#2})}%
}

\begin{document}

\title{{\Large Quantum field theory with indefinite causal structure}}
\begin{CJK*}{UTF8}{gbsn}
\author{
\vspace{5mm}
Ding Jia (贾丁)}
\email{ding.jia@uwaterloo.ca}
\affiliation{Department of Applied Mathematics, University of Waterloo, Waterloo, ON, N2L 3G1, Canada}
\affiliation{Perimeter Institute for Theoretical Physics, Waterloo, ON, N2L 2Y5, Canada
\vspace{7mm}}

\begin{abstract} 
Quantum field theory (QFT) in classical spacetime has revealed interesting and puzzling aspects about gravitational systems, in particular black hole thermodynamics and its information processing. Although quantum gravitational effects may be relevant for a better understanding of these topics, a commonly accepted framework for studying QFT with quantum gravitational effects is missing. We present a theory for studying quantum fields in the presence of quantum indefinite causal structure. This theory incorporates quantum properties of spacetime causal structure in a model independent way, and exposes universal features of quantum spacetime which are independent of the details about its microscopic degrees of freedom (strings, loops, causal sets...). 

\vspace{35mm}
\centering
\large{\textit{Essay written for the Gravity Research Foundation 2018 Awards for Essays on Gravitation.}

Submitted on March 27, 2018}
\end{abstract}

\clearpage\maketitle
\thispagestyle{empty}
 
\end{CJK*}

\clearpage
\setcounter{page}{1}

\section{Introduction}

\begin{quote}
It should not be thought unreasonable that a black hole, which is an excited state of the gravitational field, should decay quantum mechanically and that, because of quantum fluctuation of the metric, energy should be able to tunnel out of the potential well of a black hole. ... The real justification of the thermal emission is the mathematical derivation given in Section 2 for the case of an uncharged non-rotating black hole.  \cite{hawking1975particle}
\end{quote}
The above explanation of black hole radiation from Hawking's classic 1975 paper reflects a deep dilemma that troubles physicists till the present day. On the one hand quantum effects of spacetime may well provide the crucial ingredient of understanding black hole thermodynamics, but on the other hand when it comes to concrete derivations one has to descend to classical spacetime, due to the lack of a commonly accepted framework to incorporate quantum gravitational effects. 

In this essay we sketch (the details will appear elsewhere) a theory of quantum fields that incorporates quantum gravitational effects in a model independent way. Unlike string theory, loop quantum gravity, causal set theory etc. that posit new postulates about the microscopic structure of quantum spacetime, the theory presented here does not invoke any assumption beyond what is already contained in general relativity and quantum theory. In general relativity, spacetime causal structure is dynamical. In quantum theory, dynamical degrees of freedom subject to quantum uncertainties. Quantum gravity as a theory that combines quantum theory and general relativity is thus expected to have indefinite causal structure \cite{hardy_probability_2005, hardy2007towards}. The theory we present incorporates indefinite causal structure into quantum field theory (QFT).

The basic idea is to discard an underlying classical spacetime and encode the information of spacetime causal structure in the field correlations.\footnote{This idea was successfully pursued in \cite{saravani2016spacetime} for spacetime with definite causal structure.} Inspired by similar constructions in general probabilistic theories \cite{hardy_probability_2005, hardy2007towards} and operational quantum theory \cite{oreshkov2012quantum, oreshkov2016operational, chiribella2013quantum}, we generalize the traditional definition of states so that they reflect indefinite causal structure. TABLE \ref{tab:compare} summarizes the new structures of the theory.

\begin{table}[h]
\centering
\caption{Comparison of the old and new theories}
\label{tab:compare}
\begin{tabular}{|l|l|l|}
\hline
               & Ordinary QFT                                   & QFT with indefinite causal structure                                                              \\ \hline
Hilbert space  & Global space $\mathcal{H}$                     & Tensor product space $\otimes_{i}\mathcal{H}^{i}$ \\ \hline
Field operator & $\phi(x)$ in Heisenberg or interaction picture & $\phi_S(\vec{x})$ in Schr{\"o}dinger picture  \\ \hline
Spacetime & Spacetime manifold with coordinate $x$ & Causal relation inferred from correlations                  \\ \hline
State          & $\omega:\mathcal{A}\rightarrow \mathbb{C}; \phi(x)\mapsto \ev{\phi(x)}_\omega$    & $w:\mathcal{A}\times \mathcal{A}\rightarrow \mathbb{C}; (\phi_S(\vec{x}),\psi_S(\vec{x}))\mapsto \ev{\phi(x)}_\omega\overline{\ev{\psi^*(x)}}_\omega$                                                                                                 \\ \hline
\end{tabular}
\end{table}

\section{QFT with indefinite causal structure}

Ordinary QFT assumes that there is a global time foliation of spacetime with a global time evolution (usually taken to be unitary). This foliation implies a global definite causal order, so cannot be retained in a theory with indefinite causal structure. With the global foliation and evolution discarded, the new theory calls for the following changes.
\begin{itemize}
\item Assign local Hilbert spaces to each place where field operators apply.
\item Use field operators in the Schr{\"o}dinger picture.
\item Update $x$ or $\vec{x}$ from a coordinate on a spacetime manifold to a label of local Hilbert spaces.
\item Define states as linear functionals that map to ``squares'' of the correlation functions.
\end{itemize}

\textbf{Tensor product spaces.} In ordinary QFT, a global time foliation makes it possible to work with a single global Hilbert space. In the new theory without a global time foliation, each place where field operators apply gets assigned a local Hilbert spaces. The tensor product of the local spaces forms the global space. Note that the causal relations among the regions can be anything, in contrast to traditional conceptions of quantum theory for which only spacelike separated systems form tensor products. This means that, for example, even the same system at two different times gets assigned two distinct Hilbert spaces. The reason for this move is to incorporate indefinite causal structure. We allow tensor product formation for arbitrary local systems because their causal relation may be a ``quantum superpotition'' of causal and acausal.

\textbf{Schr{\"o}dinger picture field operators.} The global time foliation and evolution of ordinary QFT also show up in the definition of the field operators with time dependence. In the usually adopted Heisenberg or interaction picture a field operator $\phi(x)$ has a global time evolution built into it. For example, the Heisenberg picture operator $\phi_H(t,\vec{x})$ relates to the Schr{\"o}dinger picture operator $\phi_S(\vec{x})$ through $\phi_H(t,\vec{x})=U^*(t)\phi_S(\vec{x})U(t)$, where $U(t)$ is a unitary time evolution from the reference time of $\phi_S(\vec{x})$ to the time $t$ and star denotes adjoint. 

The new theory adopts the Schr{\"o}dinger picture field operators $\phi_S(\vec{x})$, because due to indefinite causal structure a global time evolution does not exist (it would imply a fixed causal order for the global system at different times). The information propagation properties of the systems will instead be encoded in the generalized state.

\textbf{$x$ or $\vec{x}$ as labels for local spaces.} Events in a classical spacetime manifold have a definite causal structure so cannot be retained. In the new theory the spacetime causal structure is to be inferred from the field correlations, and the argument $x$ or $\vec{x}$ of the field $\phi$ becomes a label of the local spaces where $\phi$ applies.

\textbf{States for the ``square'' of the correlation functions.} In ordinary QFT, a state is a linear functional from the algebra of field operators $\mathcal{A}$ to the value of the correlation function. 
If $\phi(x)\in \mathcal{A}$ and $\phi(y)\in \mathcal{A}$ are two Heisenberg picture field operators, a state $\omega$ maps $\phi(x)\phi(y)\in \mathcal{A}$ to the two-point function $\ev{\phi(x)\phi(y)}_\omega$. Similarly $\phi(y)\phi(x)\in \mathcal{A}$ gets mapped to $\ev{\phi(y)\phi(x)}_\omega$. A non-zero commutator $\ev{[\phi(x),\phi(y)]}_\omega:=\ev{\phi(x)\phi(y)}_\omega-\ev{\phi(y)\phi(x)}_\omega$ indicates that $x$ and $y$ are causally related.

In the new theory with tensor product spaces and Schr{\"o}dinger picture field operators, the corresponding operators are $\phi_S(\vec{x})\otimes \id(\bar{x})$ and $\phi_S(\vec{y})\otimes \id(\bar{y})$, where $\id(\bar{x})$ and $\id(\bar{y})$ are the identity operators on the Hilbert spaces complement to $x$'s and $y$'s. These operators belong to the algebra $\mathcal{A}'$ of operators on the global tensor product space. Since $\phi_S(\vec{x})$ and $\phi_S(\vec{y})$ act on different subsystems, $[\phi_S(\vec{x})\otimes \id(\bar{x}),\phi_S(\vec{y})\otimes \id(\bar{y})]=0$. A state $\omega$ defined as a linear functional from $\mathcal{A}'$ to the correlation function $\ev{[\phi(x),\phi(y)]}_\omega$ would always vanish. This would incur an issue for recovering ordinary QFT as a special case in the new theory.

The resolution is to let the state map to the ``square'' of the correlation functions. Suppose $w$ is a generalized state in the new theory that corresponds to the state $\omega$ in the old theory. Consider
\begin{align*}
w:\mathcal{A}\rightarrow \mathbb{R};\quad \phi_S(\vec{x})\mapsto\abs{\ev{\phi(x)}_\omega}^2,
\end{align*}
where starting here for simplicity we omit identity operators and write $\phi_S(\vec{x})$ for $\phi_S(\vec{x})\otimes \id(\bar{x})$. Since for the squared correlation the ``commutator'' \textit{should} always vanish, i.e., $\abs{\ev{\phi(x)\phi(y)}_\omega}^2-\abs{\ev{\phi(y)\phi(x)}_\omega}^2=0$, the conflict is resolved.

To obtain ordinary QFT as a special case, one needs to recover $\ev{\phi(x)}_\omega$. For this propose extend $w$ to (bar denoting complex conjugate and star denoting adjoint)
\begin{align*}
w:\mathcal{A}\times \mathcal{A}\rightarrow \mathbb{C};\quad (\phi_S(\vec{x}),\psi_S(\vec{x}))\mapsto \ev{\phi(x)}_\omega\overline{\ev{\psi^*(x)}}_\omega.
\end{align*}
This is an extension because by setting $\psi_S(\vec{x})=\phi_S^*(\vec{x})$ one recovers $\abs{\ev{\phi(x)}_\omega}^2$. Now by setting $\psi_S(\vec{x}')=\id$ one recovers $\ev{\phi(x)}_\omega$.  Here $\vec{x}$ and $x$ can label product systems that contain multiple subsystems. The two-point functions are recovered as $\ev{\phi(x)\phi(y)}_\omega=w(\phi_S(\vec{x})\otimes \phi_S(\vec{y}),\id)$ and $\ev{\phi(y)\phi(x)}_\omega=\overline{\ev{\phi(x)\phi(y)}}_\omega=w(\id,\phi_S^*(\vec{x})\otimes \phi_S^*(\vec{y}))$, from which the commutator and anticommutator can be obtained. An example of a generalized state $w$ is presented below for illustration.

\begin{figure}
    \centering
    \includegraphics[width=.6\textwidth]{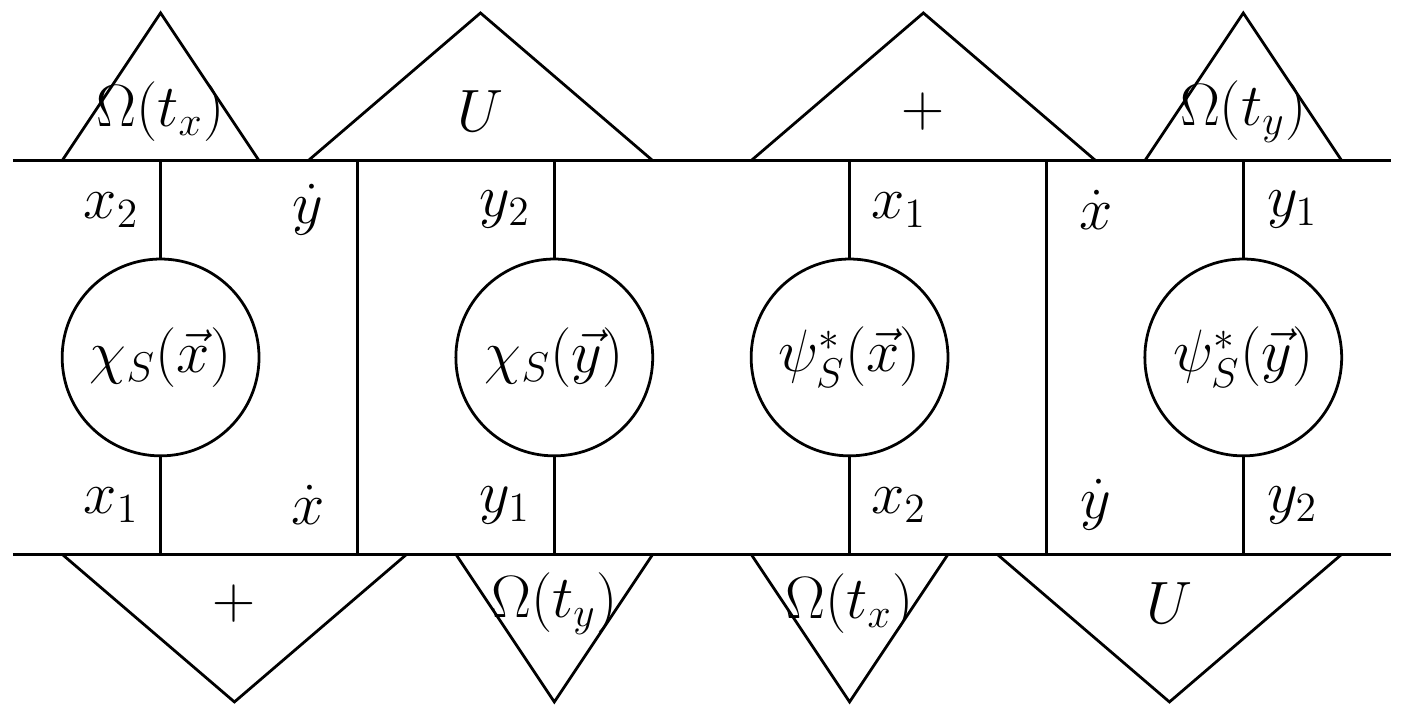}
    \caption{Illustration for recovering two-point function}
    \label{fig:my_label}
\end{figure}

\textbf{Representation of the generalized state as a vector.} In ordinary QFT the two-point function in the Schr{\"o}dinger picture reads $\bra{\Omega(t_x)}\phi_S(\vec{x})U(t_x-t_y)\phi_S(\vec{y})\ket{\Omega(t_y)}$ for the state $\omega$ corresponding to $\ket{\Omega}$. The corresponding state $w$ has a vector\footnote{$\ket{W}$ is actually not a vector in the Hilbert space because it is unnormalized. Rigorously one can use the GNS construction for $w$ to find the proper vector representation, but we don't delve into this topic here.} representation $\ket{W}=\ket{+}^{x_1\dot{x}}\ket{\Omega(t_y)}^{y_1}\ket{\Omega(t_x)}^{x_2}\ket{U}^{\dot{y}y_2}$, where $\ket{+}^{x_1\dot{x}}=\sum_i \ket{ii}^{x_1\dot{x}}$ with $\{\ket{i}\}$ a set of orthonormal basis is an unnormalized vector, and $\ket{U}^{\dot{y}y_2}=U(t_x-t_y)\ket{+}^{\dot{y}y_2}$ with $U(t_x-t_y)$ acting on $\dot{y}$.\footnote{This setup is reminiscent of the multiple time formalism \cite{aharonov2009multiple}. The difference is that here the generalized state maps to the correlation function and no post-selection is invoked.} The two-point function can be recovered as follows (FIG. 1). Start with
\begin{align*}
\bra{W}\chi_S(\vec{x})\chi_S(\vec{y})\psi^*_S(\vec{x})\psi^*_S(\vec{y})\ket{W}=\bra{W}\chi_S(\vec{x})_{x_1}^{x_2}\otimes \chi_S(\vec{y})_{y_1}^{y_2} \otimes \id_{\dot{x}}^{\dot{y}}\otimes\psi^*_S(\vec{x})_{x_2}^{x_1}\otimes \psi^*_S(\vec{y})_{y_2}^{y_1}\otimes \id_{\dot{y}}^{\dot{x}}\ket{W},
\end{align*}
where subscripts (superscripts) label the domain (image) spaces. By the definition of $\ket{W}$, setting $\chi=\phi$ and $\psi=\id$ yields $\bra{\Omega(t_x)}\phi_S(\vec{x})U(t_x-t_y)\phi_S(\vec{y})\ket{\Omega(t_y)}=\ev{\phi(x)\phi(y)}_\omega$, while setting $\chi=\id$ and $\psi=\phi$ yields $\bra{\Omega(t_y)}\phi^*_S(\vec{y})U^*(t_x-t_y)\phi^*_S(\vec{x})\ket{\Omega(t_x)}=\ev{\phi(y)\phi(x)}_\omega$.

\textbf{Indefinite causal structure.} Now we give an example of a state that incorporates indefinite causal structure. Suppose $\ket{W}$ above describes the situation that $y$ causally precedes $x$, and denote $\ket{W}$ as $\ket{W_{y\rightarrow x}}$. Recall that the new theory has no underlying classical spacetime, so the causal relation is inferred from the vector $\ket{W_{y\rightarrow x}}$ itself (e.g., from $\ket{U}$ defined with respect to $U(t_x-t_y)$ that actually propagates causal influences from $\phi_S(\vec{x})$ to $\phi_S(\vec{y})$). Without the burden of a fixed causal relation between $x$ and $y$, we can define $\ket{W_{x\rightarrow y}}=\ket{\Omega(t_x)}^{x_1}\ket{+}^{\dot{y}y_1}\ket{V}^{x_2\dot{x}}\ket{\Omega(t_y)}^{y_2}$, which propagates causal influences in the opposite direction through another unitary $V$. Then the new state
\begin{align}
\ket{W}=\frac{1}{\sqrt{2}}\ket{W_{x\rightarrow y}}+\frac{1}{\sqrt{2}}\ket{W_{y\rightarrow x}}
\end{align}
encodes a ``superposition'' of $y$ causally preceding $x$ and $x$ causally preceding $y$. 

\section{New physics?}

What new features can be expected from the new theory? There have been some interesting suggestions from analyzing finite dimensional models. Firstly, we expect that the quantum gravitational fluctuations of the spacetime causal structure regularize the UV divergence of the correlation functions \cite{jia2017quantum, jia2017generalizing}. Technically this is a consequence of the strong subadditivity of the von Neumann entropy. The intuition behind this regularization mechanism is that causal fluctuations spread out the correlation between $x$ and $y$ to ``correlation channels'' with different causal relations and thereby reduce the strength of the correlation. 

Secondly, we expect that information can leak out of black holes in quantum spacetime without violating causality (without superluminal signalling) \cite{jia2017quantum, jia2018analogue}. The communication capacity induced by causal fluctuations can be calculated and even the tiniest causal fluctuation generically induce a positive capacity. The definitions of black holes in theories of classical spacetime as containing strict causal barriers is then best viewed as an approximation. There is no \textit{superluminal} signalling because light can still reach wherever other forms of matter reach.

Both the divergence of the correlation function (through the Hadamard condition)\footnote{C.f. \cite{agullo2009insensitivity, helfer2010comment, agullo2010reply, nicolini2011minimal}.} and the existence of a causal barrier can affect black hole thermodynamics and its information loss problem.\footnote{It has been argued that the causal barrier of an event horizon is not essential for Hawking radiation, but even if so the barrier is still essential for the information problem \cite{barcelo2011minimal}.} Therefore the new theory of QFT with indefinite causal structure has the potential of establishing results that are unexpected or only speculated by ordinary QFT in classical spacetime.

\section*{Acknowledgement}
I am very grateful to Lucien Hardy and Achim Kempf for guidance and support as supervisors, and to Jason Pye and Vern Paulsen for valuable discussions.

Research at Perimeter Institute is supported by the Government of Canada through the Department of Innovation, Science and Economic Development Canada and by the Province of Ontario through the Ministry of Research, Innovation and Science.  This work is supported by a grant from the John Templeton Foundation. The opinions expressed in this work are those of the author's and do not necessarily reflect the views of the John Templeton Foundation.

\bibliographystyle{apsrev}
\bibliography{bib.bib}

\end{document}